# WHY MASS MEDIA MATTER TO PLANNING RESEARCH:
## THE CASE OF MEGAPROJECTS†


By Bent Flyvbjerg, University of Oxford





ABSTRACT

This article asks how planning scholarship may effectively gain impact in planning practice through media exposure. In liberal democracies the public sphere is dominated by mass media. Therefore, working with such media is a prerequisite for effective public impact of planning research. Using the example of megaproject planning, it is illustrated how so-called "phronetic planning research," which explicitly incorporates in its methodology active and strategic collaboration with media, may be helpful in generating change in planning practice via the public sphere. Main lessons learned are: (1) Working with mass media is an extremely cost-effective way to increase the impact of planning scholarship on practice; (2) Recent developments in information technology and social media have made impact via mass media even more effective; (3) Research on "tension points," i.e., points of potential conflict, are particularly interesting to media and the public, and are especially likely to generate change in practice; and (4) Tension points bite back; planning researchers should be prepared for, but not afraid of, this.

† The author wishes to thank the editor of *JPER* and three anonymous reviewers for their highly useful comments on an earlier version of the article.




INTRODUCTION

The purpose of the present article is to demonstrate how planning scholars may actively and

strategically engage with mass media to have their research impact public deliberation, policy, and

practice. Flyvbjerg (2002, 2004) presents the theory and method of phronetic planning research,

which is an approach to the study of values and power in planning based on a contemporary

interpretation of Aristotelian phronesis, variously translated as practical wisdom or judgment. The

aim of phronetic research is to inform public deliberation and practice. Such research is focused on

the following four value-rational questions, asked for specific instances of planning practice in a

particular context: (1) Where are we going with planning? (2) Who gains and who loses, by which

mechanisms of power? (3) Is this development desirable? (4) What should be done, if anything?

Communicating research results to the public and to practice is an integral part of phronetic

planning research. Even in Aristotle's original definition of phronesis, laid down more than two

millennia ago, the knowledge-action relationship is clear. A defining characteristic of phronesis is,

in Aristotle's words, that it is "reason capable of action" (The Ethics, Book VI, 1140a24-1140b12).

Phronetic research results ("reason") are therefore results only to the extent they have an impact on

practice ("action"). In public affairs, reason is made capable of action by effectively having reason

enter the public sphere and public deliberation. It is reason times exposure in the public sphere that

matters, not reason alone.

Today, mass media dominate the public sphere in liberal democracies. The relationship of research

with media must therefore be reflected in attempts to better understand how research gains public

impact. However, to my knowledge no study exists that describes in detail this aspect of planning

research, phronetic or other. Even in the social sciences as a whole, studies of how scholars work

with mass media to secure public impact of their research seem rare (Bagdikian 2004, Bryant and



Oliver 2009, McCombs 2004). The article therefore asks and answers the question of how planning research may gain impact in public deliberation and practice through media exposure. Because the field is underexplored, it is found prudent to start with a phenomenological narrative case study. The study shows how a group of phronetic planning researchers, including the author, engaged with media, including world-leading ones like The New York Times, in order to place their research results on the public agenda and initiate change in their chosen field of interest: megaproject policy and planning.[1] The study ends by encouraging planning scholars to take responsibility for gaining impact with their research in public deliberation and practice by engaging with mass media.

The study is not about the broad range of ways in which planners engage with mass media. It is also not about how planning scholars may serve as media pundits, offering their opinion and commentary on many different aspects of their subject area, not necessarily based on their own research. The study is specifically about how planning scholars may gain impact on planning practice via mass media with their own publications and research results.

ZERO PUBLIC EXPOSURE EQUALS ZERO PUBLIC IMPACT

My first experiment with phronetic planning research was like throwing a stick of dynamite. What I had read by Machiavelli, Nietzsche, and Foucault proved to be true: you can effectively transform specific social and political practices by writing what Machiavelli calls the verita effettuale (effective truth) and Nietzsche wirkliche Historie (real history) of such practices. I published my carefully researched real history of corrupt city planning in my hometown, Aalborg, Denmark. This created a media storm. And this stopped the planning practices I had studied dead in their tracks, to be replaced by a more democratic and modern type of city planning. The latter was designed along lines I had proposed, and it won the European Union's "European Planning Prize" for innovative and democratic urban policy and planning, carrying the day in Brussels over 300 nominees



(Flyvbjerg 1998a; 2001: 141-61; 2002; Hansen 2002, 2006). The experience was almost too good to be true. If this was how phronetic planning research worked I wanted to do more. I also wanted to test whether the success in Aalborg was just beginner's luck.

Without the media exposure, the Aalborg study would not have had the impact on practice it did. To be sure, the exposure did not happen by chance. My university and I helped it along with press releases, a news conference, and by being available to media for interviews. But the fact that the media picked up the story across the board, locally and nationally, was the single most important factor in making change happen. The media magnified thousandfold the transparency created by the study and thus brought the problems uncovered by it to the attention of voters and the general public. Voting is taken seriously in Denmark, with an 80-90 percent turnout at most elections, so public opinion matters to officials. Confronted with a study and a strong public opinion that depicted officials and politicians as ineffective and corrupt, those responsible quickly decided to change their ways. Transparency worked.

This is not to say media were the only game changer in Aalborg. Other parts of civil society were important as well, as was government. But in contemporary liberal democracies mass media play a key role, because more than any other factor they structure and dominate the public sphere, for better and for worse (Calhoun 1992, Croteau and Hoynes 2006). Today's mass media are seriously called the "fourth power" of government, expanding on Montesquieu's three classic powers, the legislative, the executive, and the judicial. Therefore, concerned citizens and organizations bring their issues to the attention of media in order to be heard and to affect change. So do concerned phronetic planning researchers.

The main means of communication for planning academics is the scholarly text. The impact of a scholarly text in public debate depends on both its publication and exposure, apart from content.



Without publication there can be no impact. But equally, with zero public exposure there will also be zero public impact. Most scholars are concerned with academic exposure mainly and typically do not focus on public impact. Their focus is on getting published and being cited in academic media, because the culture and incentives of academic institutions encourage this. Thus it is unsurprising that most academic texts have little impact in the public sphere.[2] In contrast, phronetic planning researchers are explicitly concerned about public exposure, because they see it as a main vehicle for the type of social and political action that is at the heart of phronesis.

GAINING IMPACT THROUGH TENSION POINTS

After the Aalborg study, I decided to start research on megaproject policy and planning. Results from this research are published in Flyvbjerg, Holm, and Buhl (2002, 2005), Flyvbjerg, Bruzelius, and Rothengatter (2003), and Flyvbjerg (2007a, 2008). These publications focus on cost overruns, benefit shortfalls, risk, optimism, and deception in very large and very expensive infrastructure developments. The idea for this research grew from developments in Denmark's first megaproject, the so-called Great Belt fixed link (1987-1998), which connects East and West Denmark and links Scandinavia with continental Europe across the entrance to the Baltic Sea. The Great Belt link consists of three projects: the second-longest underwater rail tunnel in Europe, after the Channel tunnel, and two bridges, one of which was the world's longest suspension bridge at the time of completion. With a cost seventeen times that of the largest infrastructure project built in Denmark until then, the Great Belt link was a giant leap in scale for planning and construction. In effect, the Great Belt link signified the arrival of the age of megaprojects in Denmark.

But everything seemed to go wrong with the new project from the start. Costs quickly began to soar, even before construction commenced, and public opinion was massively against the venture. One of several disasters that hit the project was flooding with sea-water of the near-completed rail



tunnel due to an error in managing the tunnel boring process. A leading engineer on the project later told me that damage to the tunnel was so complete it would have been cheaper to bore a new tunnel than to rescue the existing one. However, the loss of face this would entail – leaving an unused, flooded, billion-dollar tunnel sitting permanently in the seabed as a negative monument to politics and engineering – was unacceptable to those in charge. So the flooded tunnel was emptied out, repaired, and boring completed. And, sure enough, when the tunnel finally opened to traffic the cost overrun was 120 percent in real terms and it proved financially non-viable. The government had paid for two tunnels but got only one, with revenues from the tunnel insufficient to cover the escalated costs.

The events at Great Belt got me interested in answering the following specific research question: "Was Denmark just unfortunate with the planning and implementation of the Great Belt megaproject, as its promoters argued, or are cost overruns and disasters like those observed for this project common for megaprojects in general?" No one had systematically answered this question, despite the fact that worldwide hundreds of similar projects had been planned and built and trillions of dollars spent. A few case studies existed and studies of small samples of projects (Hall 1980, Morris and Hough 1987, Fouracre et al. 1990, Pickrell 1990, Kain 1990, Wachs 1990). But not a large-sample study that would give a valid and reliable picture of performance in this costly field of policy and planning. I decided to do such a study to document the verita effettuale of megaprojects. To fund the research, I applied for, and was awarded, a large grant.

Then something really interesting happened. I was threatened by the highest-ranking government official in infrastructure planning in Denmark, who told me – face-to-face and in no uncertain terms – that if my studies produced results that reflected badly on his government and ministry he would personally make sure my research funds dried up. Most likely, the threat was triggered by critique I had raised against management at the Great Belt project for misinforming Parliament and the public



about cost overruns on the project. Management said costs had overrun by 29 percent total, for the tunnel and two bridges combined. My research showed the figure to be 55 percent, a difference of several billion kroner. Every time management would publicly mention their 29 percent figure, I would just as publicly point out that the real overrun was 55 percent. This happened over and over until management did an opinion poll asking the public which of the two figures they thought was right. The public said 55 percent, and this made management give up on postulating their 29 percent and acknowledge my 55, a figure later verified by the National Audit Office of Denmark (1998). For a scholar, a poll seems a strange way to decide the truth value of data, but it taught me the lesson that it is possible to inject research results into and influence the often highly manipulated public discourse on megaprojects. It also taught me that blowing the whistle repeatedly on the management of the largest and most prestigious planning project in Danish history had probably miffed people in high places. More importantly, the threat indicated to my colleagues and me that we may have hit pay dirt even before properly beginning the research. If the Danish government found it worth their while to dispatch their Chief Planner to threaten me over a piece of research, this was the best demonstration we could get that the research must be done and was likely to produce interesting results.

The Chief Planner inadvertently alerted us to the type of power relation we call "tension points," similar to Foucault's "virtual fractures." These are "lines of fragility in the present ... which open up the space of freedom understood as a space of concrete freedom, that is, of possible transformation" (Foucault 1998: 449-450). This type of power relation is particularly susceptible to problematization and thus to change, because it is fraught with dubious practices, contestable knowledge, and potential conflict. Thus even a small challenge – like problematization by scholars – may tip the scales and trigger change in a tension point. Problematizing tension points may be compared to hitting a rock with a hammer. If you hit the rock at random it seems unbreakable, even if you hit it hard. If you strategically hit the rock at the small, near-invisible fractures that most



rocks have, the rock will break, even if you hit it gently. Tension points are the fractures that phronetic planning researchers seek out; this is where researchers hit existing practices to make them come apart and create space for new and better ones.

Tension points have the added advantage that with their focus on power and dubious practices they tend to make good stories and thus to be of interest to media and the public. Flyvbjerg, Landman, and Schram (2012) expands on the use of tension points in phronetic research. Here it is enough to say that the Chief Planner made us aware of the acute tension between the <u>image</u> that he and the government wanted to project for megaproject planning and the <u>reality</u> of such planning. This tension point proved an important lead for our research and got us off to an effective start in problematizing megaproject policy and planning.[3]

THE POWER OF KNOWLEDGE

While my colleagues and I were busy collecting data on megaprojects from around the world, the Danish government approved two more such projects in addition to the Great Belt link. The new projects were the Øresund bridge between Denmark and Sweden (1991-2000), the world's longest combined cable-stayed bridge for road and rail, and the Copenhagen metro (1992-2007), one of the first driverless metros in Europe. We decided to test our methodology, described in Flyvbjerg (2002, 2004), on the three Danish projects before going global. Both the Øresund bridge and the Copenhagen metro were under construction, and like the Great Belt link the two new projects quickly ran up billion-dollar cost overruns that threatened their viability.

The project promoters, including the Danish government, had now made the same error three times over on three different megaprojects. They had underestimated costs by billions of dollars and thereby placed large sums of citizens' money at risk. We pointed this out in public. We also told



Parliament, who approve megaprojects in Denmark, and citizens, who underwrite projects financially, that they might want to consider the thought that an error made three times straight was perhaps not a random error. Maybe the "error" was better understood as either incompetence, that is, not learning from experience, or intentional misrepresentation, that is, lying, the latter being unlawful in Denmark for this type of situation.

The public and the media quickly got interested in our work and since then we have never lacked outlets for or interest in our problematizations of Danish megaproject policy and planning, including on national TV, radio, newspapers, and the web, plus endless invitations to lecture on our research. Literally hundreds of newspaper articles have covered and quoted the research in Denmark, as have dozens of TV and radio programs – some of them with hour-long coverage of the work – on both national and local media. We spend time working with the media and the public, because phronesis is aimed at public debate and in modern society public debate takes place in the media to a large extent. As a result we have made it more difficult for promoters to have the delusions and deceptions accepted that are at the core of conventional megaproject policy and planning (Flyvbjerg, Garbuio, and Lovallo 2009).

While the public and the media liked our work, this was not the case for the powers that be. We had actively generated the type of situation the government's Chief Planner had wanted to avoid when he warned me to not do research that reflected badly on his government. In fact, as we gained momentum with our critique, I was called on the carpet in Copenhagen by the Chief Planner's boss, the Minister of Transportation. At the meeting, the Minister complained that our research was making front-page news, placing the Ministry in a bad light. The Minister also assured me that the repeated errors in the budgets for the Danish megaprojects were honest mistakes; the Minister's own planners had told the Minister so. Maybe the meeting was designed to intimidate my colleagues and me by reminding us that those in power were watching us, but my main take-away was, again, that



if our work got this kind of top-level attention from government it must be because we were on to something and because our problematizations were beginning to work.

The megaproject promoters tried to undermine us. When we problematized their actions they would try to problematize us right back. Each of the government-owned multi-billion-dollar public corporations that were running the three Danish megaprojects had their own well-staffed PR offices. They, with their top brass, got on our case. This is fair enough, of course, as critique is what makes scholarship and democracy work. Except sometimes the PR people's critique would be as deceptive as the cost-benefit analyses used to justify their projects. That happened, for instance, when a former mayor of Copenhagen, who had recently become CEO of the Copenhagen metro through a controversial political appointment, tried to publicly label me as "anti-rail" and as generally "holding animus" towards the metro. He did this despite the fact that I was on record stating that a metro was good for Copenhagen and more metro should be built, just not in the mismanaged way it was now done.

Top management at the metro got particularly agitated when – during a week where Lego, the Danish toy maker, and the Copenhagen metro both discovered billion-kroner holes in their budgets due to mismanagement – I pointed out on national TV that the CEO of Lego had been sacked, whereas the metro CEO again got an extra billion and stayed on, and perhaps this was why mismanagement at the metro continued year after year. The fact that the metro CEO had to resort to name-calling, like "anti rail" for someone who supports rail, showed to my colleagues and me, and to many others, that the metro leadership had no real arguments to counter our critique. This strengthened both the critique and our predictions that without changes the metro would most likely end up as financially non-viable (Flyvbjerg 2007b: 19-24). In fact, the Copenhagen metro later became insolvent and was saved only through further subsidies and financial restructuring. Lego made a spectacular turnaround with a new CEO.



More problematically, the PR people on the megaprojects would not fight fairly and in public. Out of the public eye, some of them would routinely contact journalists and editors who covered our research and try to undermine our credibility. This is one of many ways in which power tries to subvert knowledge that does not suit its purposes. And if you do not counter as a researcher, knowledge will be weak (Flyvbjerg 1998a: 227-34). Fortunately for us, many journalists and editors were on our side and tipped us off. Serious journalism and serious research share and work for the same basic values, namely truth and democracy.[4] Both tend to see PR people as working in the less dignified world of spin, which we cannot help but see as a target to expose .

So when media were contacted by PR people who tried to put a negative spin on our research, the media would often alert us. The PR people were clearly peeved when a "personal" email they had sent to a journalist or editor stating that our work was faulty had been forwarded to us and now reappeared in the PR people's inbox with our request for them to kindly document the postulated faults. When they could not do this, which was every time, we would report it back to the journalists who had tipped us off and they would have another news item. This went on for years, and still does. It is a practice that forms part of Foucauldian micropolitics and it is work that researchers need to do to protect the power of knowledge and to make debate transparent. If the work is not done, knowledge and public deliberation suffers. I believe it has gained us much credibility with media and the public that we are willing to chase misinformation about megaprojects and abuse of power wherever they appear, even in "personal" emails to journalists and editors from promoters of megaprojects.



THE IDEAL PLANNING SCHOLAR

Our research seemed to strike a chord with the general public. People were tired of the repeated cost

overruns, benefit shortfalls, and the endless excuses and false promises made by project promoters

and their planners. To illustrate just how widespread negative public sentiments were, consider that

when a popular Danish financial magazine asked its readers to choose the most wasteful person in

Denmark among ten nominated for their misuse of citizens' money, the readers unambiguously

chose the chairman of the Copenhagen metro, a former treasurer for Denmark and vice-president of

the European Commission. The chairman was nominated for "having no control whatsoever over

costs for the Metro, as regards both operating and construction costs, not to speak of ridership."

(Penge og Privatøkonomi, 2004, 2005). The public seemed delighted that somebody was doing

research that exposed and explained this type of behavior and called for accountability.

Members of Parliament also picked up on the research. They used it for asking questions in

Parliament of those responsible for megaprojects, typically the ministers of transportation and

finance, in order to hold them accountable. Staff with the Auditor General of Denmark told us our

research had triggered decisions with the Auditor General to audit first the Great Belt link and later

the Øresund bridge and Copenhagen metro. I consulted for the Auditor General on some of these

audits and made our data available to them, so the audits could be more effective, increasing the

impact of the research.

Furthermore, I was invited by an independent government body to head an expert group that did a

year-long study on how to design the planning process for the next multi-billion-dollar project in

Denmark, the Fehmarn Belt rail and road link, which is one of the biggest infrastructure projects in

the world crossing the Baltic Sea between Denmark and Germany. The media nicknamed our

proposal "megaprojects without tears," because it explicitly tried to avoid the problems that had

beset Great Belt, Øresund, and the Copenhagen metro. The proposal was accepted by parliament



and is perhaps the clearest example of a specific megaproject in which public deliberation and the planning process was from the beginning influenced by our research, as described in Flyvbjerg, Bruzelius, and Rothengatter (2003: 143-51).

Later, the Parliament Treasury Committee and the Treasury passed government regulation that made methods mandatory that we had developed for estimating more accurately the costs and risks of major transportation projects, gaining impact for our research across the board for all such projects (Danish Ministry of Transport 2006, 2008). Finally, a new minister of transportation appointed me to a policy-proposing national infrastructure commission (Danish Infrastructure Commission 2008). These incidents are mentioned here to stress that my colleagues and I, like other scholars, had many other avenues of impact for our research than media. But media is the focus here and media exposure was crucial for getting the research known in the first place to policy makers and the public and thus for gaining practical consequence for the research results.

We found that government was not monolithic, as any Foucauldian and other true pluralists know. Government was beginning to fracture and to work against itself regarding megaprojects. By strategically doing our research on tension points, that is, the dubious practices of conventional megaproject planning, we had deliberately contributed to this fracturing. The fracturing spread to those parts of government that were responsible for developing the four Danish megaprojects. Some staff began to change sides and to act as "guerrilla employees," that is, working against the wishes of their superiors, because the employees were dissatisfied with the superiors' actions. Such staff began to see that conventional megaproject planning ran against basic values of economic efficiency, truth-telling, and due democratic process. They wanted to help remedy the situation. For strategic reasons, the guerrilla employees typically chose not to go public with their concerns. They moved clandestinely behind the scenes, leaking information from the inside, to us and others, as "salmon swimming against the current of power" (O'Leary 2010: 8).



For instance, when the Danish government began planning in earnest the Fehmarn Belt megaproject, the government's Chief Planner was so fed up with public debate, and with our problematizations, that meetings about the new project were planned to be held outside the borders of Denmark, in secret. In this manner the Danish Freedom of Information Act would not apply and nosy journalists, researchers, and citizens would be unable to find out what was going on. But unfortunately for the Chief Planner, guerrilla employees – his own colleagues – leaked his Machiavellian scheme to the media and to me, among others. A major scandal ensued, and now it was the turn of the Chief Planner's boss, the Minister of Transportation, to be called on the carpet – in Parliament – to be questioned by MPs for them to consider whether the Minister should be fired. The MPs took offense that the country's laws of freedom of information were being circumvented. The Minister apologized profusely and explained that the Chief Planner had gotten carried away and that the Minister had not known what was going on. The Minister survived, but the Chief Planner did not. After an extended period of leave he returned to a less demanding position.

Practicing phronetic planning research in the way described above can be a conflictual, and, in some instances, adversarial process. This is particularly the case as scholars identify tension points and then insert their research into this breach. For many researchers, and perhaps especially early-career scholars and graduate students, this level of public attention and conflict may be intimidating. But the question is, what is the alternative in a practice-oriented field like planning? Do we really want our young scholars to become conventional academics, concerned mainly about academic impact, that is, journal citations, instead of public, practical impact? I suggest not. To be effective, planning scholars need to be concerned about both types of impact, and the two are equally important. Planning research is different in this sense from other academic fields because of its strong orientation towards practice and the public, and this needs to be reflected in its methodology, as attempted above. Thus the ideal planning scholar, young or old, does not hide in



the ivory tower of academia; he or she has the courage and skills to enter into the fray of public debate and practice. In doing so, young scholars are best advised to get their feet wet step by step. Developing a "test lab," like I did in Aalborg, may be recommended. Here I cautiously sent up my "trial balloons," as described in Flyvbjerg (2004), to hone my methodology and ensure I made my mistakes and learned from them in a small, local context, before going national and international with the research. This – and a university that fully encourages and supports media exposure of research  – were the key ingredients in becoming confident in engaging in high-profile, public conflicts when this is necessary for impacting planning practice.

## FROM LOCAL TO GLOBAL IMPACT

While researching the Danish projects mentioned above, my colleagues and I had succeeded in collecting data on projects worldwide, eventually covering 20 nations, five continents, and seven decades. The data document that the problems with cost overruns and benefit shortfalls, first identified for Denmark, are global. So are the causes. The data basically show that megaproject promoters and their planners have to be either fools or liars to keep underestimating costs and risks, and overestimating benefits and viability, on project after project, decade after decade.

The data show nine out of ten projects have cost overruns. Overruns over 50 percent are common, over 100 percent not uncommon, as happened with Boston's Big Dig, Denver's International Airport, UK's Humber Bridge, and many other projects. Standard deviations are large, indicating risk to the second degree, that is, risk of cost overrun and risk of the overrun being much larger than expected. Most interestingly, overruns have been constant for the 70 years for which data are available, indicating that no improvements in planning and managing costs have been made over time (Flyvbjerg, Holm, and Buhl 2002; Flyvbjerg, Bruzelius, and Rothengatter 2003).



Benefit shortfalls generally mirror cost overrun. A majority of projects have benefit shortfalls. Shortfalls around 40-50 percent are common, 75 percent not uncommon, and again the standard deviations are large. Thus benefit shortfalls compound the risk that projects end up non-viable, as happened with the Channel tunnel, the Great Belt tunnel, and the Los Angeles, Miami, and Copenhagen metros. For benefits, as for costs, historical data indicate no improvements over time in estimation and planning skills (Flyvbjerg, Holm, and Buhl 2005; Flyvbjerg, Bruzelius, and Rothengatter 2003).

A distinct effect of establishing the first large dataset with valid and reliable information on cost overruns and benefit shortfalls in large-scale infrastructure projects, has been that promoters and planners can no longer, as was their habit, explain away overruns and shortfalls in specific projects as one-off instances of bad luck that are unlikely to happen again. We now know that a new project has a more than 90 percent risk of having either cost overruns, benefit shortfalls, or both. Cost overruns and benefit shortfalls are inherent to megaprojects. MPs concerned about government overspend, community groups fighting specific projects, banks whose money are at risk, government officials and national audit offices serious about curbing wasteful projects, and media who know a good story when they see one have all been quick to seize on these results for their own purposes.

All we had to do was collect the data (which was difficult enough and took several years) and communicate the results effectively to media around the world, and immediately the results got an army of feet to walk on. This alone has helped change discourse and policy for megaprojects and is an important first step in the problematization that is at the core of phronetic research. The new data have changed the power balance in megaproject policy and planning by giving ammunition to stakeholders who had little or no influence before. Planning has become less closed, undermining



the dominance of promoters and planners who had a virtual monopoly on information until our data came along.

Today it is basically impossible for promoters and planners to postulate certain costs, benefits, or risks for their projects without taking into account our data. If they try to ignore the data, other stakeholders – with an interest in exposing the real costs, benefits, and risks of projects – will bring them up, or we will. That is how policy and planning should work in pluralistic democracies.

One might argue, as did one reviewer of the present article, that to target practicing planners for producing misleading forecasts of costs and benefits is wrong, because it makes planners take the rap for the misbehaviors of politicians and elected officials. Planners are frequently powerless in the face of economic and political forces that demand development at any cost, according to this argument.[5] I will contend here that this line of reasoning is letting planners off the hook too easily. Even if planners were just capitulating to dominant economic and political forces when preparing misleading cost-benefit forecasts for megaprojects this would still entail unethical and sometimes even unlawful behavior, in many of the cases studied here. If planners want to be credible as a profession they need to deal with such unlawfulness and breaches of professional ethics, as do other professions. This is not to say that ethical planners do not exist. But as a profession planners ignore or try to dodge this dark side of planning and tend to paint instead a "sunny, relentlessly positive" image of planning, in the words of one informant. This includes, not least, the organizations that are supposed to uphold the ethics of the profession, such as the APA, RTPI, PMI, and APM. By not penalizing fraudulent forecasts, and thus quietly accepting them, professional organizations become co-responsible for their existence.[6]



A former president of APA, interviewed for the present study, provides the following explanation of the reasons why planners misinform about costs and why professional planning organizations do little about it:

> "I believe planners and consultants in general deliberately underestimate project costs because their political bosses or clients want the projects. Sometimes, to tell the truth is to risk your job or your contracts or the next contract. I think the professional organizations [like APA] are likely to simply point to their code of ethics and let it go at that [i.e., enforcing no sanctions for those who violate the code]. For example, Article B 11 in the AICP [American Institute of Certified Planners] code says: 'Planning process participants should not misrepresent facts or distort information for the purpose of achieving a desired outcome', [but it is not enforced]."[7]

As for APA, we found that not only are they doing conspicuously little to penalize planners' unethical forecasts, they are also actively suppressing the spread of information about the existence of such forecasts. Thus when we submitted our study for publication in APA's flagship academic journal, *JAPA*, APA actively tried to obstruct its distribution to media, because they saw the study as presenting planners in a bad light. This incident and its broader implications are explored in Flyvbjerg (forthcoming). It is mentioned here only to underscore two important lessons about gaining media impact for research. First, if your research results are newsworthy no one can stop you from spreading them via mass media. This is particularly true today with the internet and social media. Encouragingly, not even the biggest and most powerful professional organization in our field, APA, could stop media exposure of our research when they tried. Second, we found that you don't need PR people for getting in touch with media, not even world-class ones. Most media are inundated with PR professionals and like to deal with researchers directly, so you can cut out the middleman.



For our first international publication (Flyvbjerg, Holm, and Buhl 2002), we decided to offer exclusives to the editors of The New York Times and The Sunday Times of London. We figured that if two major newspapers like these would cover our study, then the rest of the media would follow. This proved to be correct. At the time we had no experience with major international media, but we had experience from Denmark. Here we had found that media contact for new work is simple: If you have a newsworthy story – and a good press release – you float; if not, you sink. We reckoned the same would hold true for international media. This also proved correct. The New York Times and The Sunday Times both decided to run exclusives on our study. Moreover, while the two newspapers were printing their exclusives, we emailed copies of the study and a press release written by us to every other main media in the English-speaking world, including the US. Many of these media covered the study. Distributing news like this is amazingly fast and efficient with modern information and communication technologies.

We experienced a minor, but potentially ruinous, hiccup in our contact with The New York Times. It is mentioned here to illustrate how you have to be prepared for anything, and for thinking on your feet, when dealing with media. A few days before publication, I learned that The New York Times reporter who had been assigned to write the story about our study, and who had interviewed me at length over the past weeks, had unexpectedly dropped everything and rushed to the hospital where his wife was giving premature birth to their first child. I contacted the reporter's editor, who said it was too late to place someone else on the story. He was sorry, but they had to abandon it. It was just one of those things that happen at a big, busy newspaper. He hoped I understood.

I did understand, of course, why the reporter had to drop everything. But I also understood that our strategy of gaining practical impact for our research through the public sphere would be seriously set back. We had offered an exclusive to The New York Times, so no other media were in place to



cover the story in the US. We would lose the whole country, the biggest and most important one. We could not let this happen. But what to do? Somehow the thought struck me that maybe we could play on the American fear of litigation. I got the idea from the Times editor himself. He had demonstrated this fear just a few days earlier, when he had our study checked for legal liabilities. I now explained to the editor that he had actually agreed, in writing, to run an exclusive on our study, and because of this agreement we had sacrificed opportunities with other media. If he dropped the exclusive this would incur substantial costs on us in terms of not achieving one of the key objectives of our work, namely making it public. Then I mentioned that in my view an agreement like ours was legally binding. I did not say I would sue, but I did make it clear that I thought if someone wanted to sue, they could, and that they would have a good chance of winning. I was improvising and this was clearly a long shot on my part that I feared might somehow backfire. I had no experience in dealing with media of this size and caliber, or any media in the US for that matter, and certainly not regarding legal issues. But I felt I had no choice, other than letting our strategy for going public, so central to phronesis and thus to our work, collapse. This, to me, would be equivalent to not taking our work seriously. After I made my spiel, the editor went quiet for a while. Then he said he would run the story, and he had it pieced together from the unfinished work of the reporter who had had to withdraw.[8]

The coverage of the JAPA cost study in The New York Times and The Sunday Times of London proved invaluable, both in itself and in getting the study picked up by other media. It also proved an effective first step in establishing a role for ourselves globally in informing public deliberation about megaprojects, like the one we already had in Scandinavia. In this role we would eventually help reform policy and planning in a number of countries. In the UK, Denmark, and Switzerland, use of our research and methods in project preparation has been made mandatory by government and professional organizations (Flyvbjerg and Cowi 2004; UK Department for Transport 2006a, b; Swiss Association of Road and Transportation Experts 2006). In other countries – including the US,



The Netherlands, Sweden, and South Africa – the research is being used on an ad hoc basis at the policy level and in decisions on individual projects (Transportation Research Board 2007, Dutch Commission on Infrastructure Projects 2004, Cowi et al. 2003). In 2004, we developed a method called reference class forecasting to improve project forecasts, which the American Planning Association (2005) officially endorsed; APA clearly found it easier to embrace our constructive work than our earlier problematizations:

> "APA encourages planners to use reference class forecasting in addition to traditional methods as a way to improve accuracy. The reference class forecasting method is beneficial for non-routine projects such as stadiums, museums, exhibit centers, and other local one-off projects. Planners should never rely solely on civil engineering technology as a way to generate project forecasts."

Similarly, in 2008, the administrator of the Federal Transit Administration (FTA) pointed to our research to explain how the FTA was working to increase accuracy in forecasting (Simpson 2008):

> "In his book on megaprojects and risk, Brent [sic] Flyvbjerg wrote that: 'We must find ways of institutionally embedding risk and accountability in the decision making process for mega-projects.' That's exactly where we're headed ... [A]gencies systematically over-estimate ridership forecasts, and under-estimate project costs. Until recently, we simply hadn't done enough detailed analysis to determine the reasons for this gap. And if FTA is not sure whether grantees are making accurate forecasts, then it's difficult to know whether the right investment decisions are being made."



The question of whether the FTA has succeeded in actually increasing accuracy in transit forecasting is a topic for our current research.

Finally, in 2011, we were delighted to learn that Daniel Kahneman, winner of the Nobel prize in economics for his work on decision making under uncertainty, had found reason to single out our research on reference class forecasting and how to avoid the planning fallacy in megaprojects as "the single most important piece of advice regarding how to increase accuracy in forecasting" (Kahneman 2011: 251). Most likely, Kahneman would never have heard about our work had it not been for the media coverage described above.

Working with media like this is what transforms the research from knowledge sitting in academic planning publications to knowledge that impacts practice. This is the type of knowledge that Aristotle called "reason capable of action" and Machiavelli <u>verita effettuale</u> (effective truth). Aristotle and Machiavelli both emphasized that this type of knowledge must be at the heart of social and political study if it is to matter to society. Machiavelli sought out princes to have an effect; such were his times. We seek out the public. Like princes, the public sphere is fallible and often lacks virtue. But it is key to public deliberation in modern democracies and thus inescapable for anyone who hopes to influence public decisions, including planning decisions. With our research we try to make it more difficult for the many opportunists, who invariably converge where billions of dollars and political monument building are at stake, to hijack the public debate, policy, and plans for their own purposes.

This is not to say you need media like <u>The New York Times</u> or <u>The Sunday Times</u> of London to do applied phronesis. If you are researching local issues, like I was when I did my first piece of phronetic planning research in Aalborg, and like my colleagues and I are today with individual megaprojects, then the relevant media are typically local newspapers and local TV and radio



stations. However, my colleagues and I have deliberately taken phronetic planning research from first local to national issues and then to global ones. This was done out of scholarly curiosity, as a methodological experiment, to test whether phronesis works at all three levels. We found that it does. In our experience, there is no substantive difference between doing phronetic planning research at the local, national, and global level. The only difference is scale, and the methodology scales well.

LESSONS OF MASS MEDIA AND PHRONESIS

This article may give the impression that my colleagues and I spend a lot of time on media contact. That is not the case. While we think working with media is crucial to doing phronetic planning research, and while we may spend more time than other planning researchers on media contacts it is really only a few hours or days here and there, typically weeks and months apart, but we have done it consistently over the years of doing phronesis. Working with media is intense and results are immediate and practical. As such it is a stimulating break from the slow data collection, analyses, and theory building that are at the core of academic research. Compared to the vast number of hours that go into the research, the hours spent on media are negligible; less than a percent of our total time spent on the research, albeit unevenly distributed among team members and in time, with particular pressures right after a press release has gone out.[9] The difference that this minuscule effort makes to the impact of the research on public debate, policy, and practice may be large, nevertheless, which is why the effort needs to be made, from a phronetic point of view.

My colleagues and I have learned the following from our engagements with mass media and the public sphere in doing phronetic planning research:



1. <u>In liberal democracies, working with mass media is crucial for scholarship to gain impact on practice</u>. This is because in liberal democracies the public sphere is dominated by mass media, and the public sphere is the main vehicle for scholarship to enter into public deliberation, policy, and practice. Mass media are dominated by commercial interests and raw political power, to be sure, which limits and biases access. This is a danger that needs to be addressed and managed, as pointed out by Frank (2012), but it is not a game stopper. If, as a planning scholar, you have produced research results that you can stand by and that media think are newsworthy, then you and the media have joint interests in spreading the word and gaining impact. That is how media exposure worked for the studies described above.

2. <u>Problematization of tension points is particularly well-suited for generating change and for producing research results that are of interest to the public and media</u>. Rule number one for gaining exposure and impact with phronetic planning research is to study things that matter in ways that matter. Here, problematization of social and political tension points has proved particularly effective. This is because such problematization focuses on power and on dubious practices, both of which are attractive topics to the public and to media, and both of which are often difficult to justify when exposed to the hard gaze of scholars and investigative journalists. This ensures that the combined efforts of research and journalism is likely to have an impact, which, too, is attractive to the public and to media. How to work with tension points is further described in Flyvbjerg, Landman, and Schram (2012).

3. <u>Phronetic planning research and phronetic impact is replicable across different problematics, geographies, and time periods</u>. The initial concern proved unfounded that the impact of the Aalborg study on practice was perhaps just beginner's luck. My colleagues and I were able to replicate the same type of impact with, first, our research on megaproject planning in Denmark,



and later, similar research globally. Other researchers have achieved similar impacts, as demonstrated in Flyvbjerg, Landman, and Schram (2012).

4. <u>Phronesis scales well</u>. Methodologically there is no difference between doing phronetic planning research at the local, national, and global level; the method is equally relevant and effective at each level. The only difference is scale, and the methodology scales well. For mass media, the effort it takes to get a relevant piece of research covered by <u>The New York Times</u> is approximately the same as it takes to get research covered by <u>The Aalborg Stiftstidende</u>, that is, the local paper in Aalborg that was the main vehicle for exposing the results of the Aalborg study. The economies of scale are therefore substantial regarding exposure and impact.

5. <u>Tension points bite back</u>. When you publicly problematize dubious practices and propose alternatives to replace them, do not expect a smooth ride. The people and organizations who benefit from the practices you problematize are likely to bite back. But you do not do this type of work for the smooth ride, you do it to change things. If your research triggers change, this is likely to get you both friends and enemies. If nobody is against a specific piece of phronetic planning research, most likely the research is unimportant as regards its implications for practice. Phronetic researchers are power researchers, and as such they do not expect consensus for their work, but conflict. Consensus cannot be ruled out and is desirable where possible, but a priori consensus is considered dubious, because too often it is an illusion created by disregarding power, as in Habermasian discourse ethics, or by marginalizing groups who are inconvenient to the supposed consensus (Flyvbjerg 1998b). True consensus is rare in matters of large-scale policy and planning.

6. <u>Dissemination of research results via mass media is unstoppable</u>. If your research results are newsworthy – and results of phronetic planning research often are – no one can stop you from



spreading them via mass media. This is particularly true today with the internet and social media. Not even the biggest and most powerful professional organization in our field, the American Planning Association, could stop exposure of our research results, when they tried. This is encouraging. When someone tries to stop you, guerilla employees may come in handy, and they are found in most organizations.

7. <u>Press releases matter</u>. Media contact for research is simple: If you have a newsworthy story – and a good press release – you float; if not, you sink. We found that if you are not in control of the press releases about your work, you are not in control of its possible impact on public debate. We also found that it is a bad idea to make yourself dependent on PR people for writing and disseminating press releases. It is better to write your own releases, with feedback from professional journalists, and to distribute the releases yourself. Most media are inundated with PR professionals and like to deal with researchers directly, so you can cut out the middleman. Press releases are more effective for communicating research results than are op-eds and the like. Generally, it works better for scholars to help journalists do their work by feeding journalists important and newsworthy research results than it is for scholars to try to do the work of journalists by writing journalistic articles about their own research.

8. <u>Working with mass media is an extremely cost-effective way to increase the impact of research on policy and planning, and thus to meet the phronetic imperative that research should make a difference to action</u>. Media contact takes little time, compared to the time it takes to do research; less than one percent of the total time involved, in our estimate. The difference that this minuscule effort makes to the impact of the research on public deliberation, policy, and practice is large, which is why the effort needs to be made.



9.  <u>Recent developments in information technology have made phronetic planning research even more effective</u>. For phronetic research to have an impact, it must reach people. With the internet and social media such reach has grown dramatically. The instruments for problematization and for opening up alternative courses of action that are the key results of phronetic research may be uploaded to the internet and social media, to be downloaded and used by anyone with a computer or smartphone.

10. <u>Research results take on a life of their own</u>. After being published and exposed via mass media and the internet, results from phronetic planning research take on their own life as instruments in the struggle for a better life for specific communities and individuals. This is an effect that phronetic planning researchers deliberately aim for.

I suspect many planning researchers would prefer not to have to deal with media in the manner described above and not to have to jump into the fray of public debate with their research results, risking public critique and sometimes even attempts at vilification. When it comes to media, many planning researchers seem to be wisely living according to the ancient Latin motto "<u>bene vixit qui bene latuit</u>" (they who live unnoticed live well). I also suspect this is a main reason why planning research matters so little in society today.

In contrast, imagine that a majority of planning researchers decided to (1) do research that matters to the communities in which we live <u>and</u> (2) make sure that the research results are effectively communicated to the public sphere, to be used in public deliberation, policy, and practice in those communities. Planning research and society would both be transformed for the better, where "better" is defined by the conventional values of liberal democracies, for example, better informed and more just. Moreover, planning research would stand a chance of ridding itself of its role as the poor cousin of the academy – a research field mainly ignored by other fields – and becoming a



leader in a skill that today is much sought-after across the academy and that planning research is particularly well-positioned to deliver: how to effectively translate research results into action.

This article shows that such an alternative course of action for planning research is possible and can be replicated and scaled. The question is not whether it can be done. The question is whether we planning researchers want to do it or not. I suggest we do, in order to strengthen truth and democracy in the societies where we live and, not least, to strengthen planning research itself.

NOTES

---

[1] Megaprojects are multi-billion-dollar public infrastructure projects, each with the potential to transform cities, regions, and the lives of millions. Ever more and larger megaprojects are currently being built around the world, in what <u>The Economist</u> (June 7, 2008: 80) has called the "biggest investment boom in history." Enormous amounts of money are at stake in potential waste of citizens' money as are massive social and environmental impacts and due democratic process.

[2] By a combined use of Google Scholar (or the ISI Web of Knowledge) and Google News, it is easy to establish that most academic publications have little or no impact in both mass media and academic media. Publications may have other important impacts, needless to say, in teaching students for example. Also, scholars may have media impact that is not directly related to the publication of their research results, for instance as pundits who offer their opinion or commentary on many different aspects of their subject area, not necessarily based on their own research.

[3] Political action requires constructively getting things started again, and this is the second step in phronetic planning research, after problematization. As planning researchers, my colleagues and I problematize and deconstruct



megaproject policy and planning. As planners, we constructively design proposals for improvements to what we problematize. Thus we are not satisfied with only writing the *wirkliche Historie* of what we study and with problematizing it, although we see problematization and critique as the most important drivers of progress. Examples of our constructive work may be found in Flyvbjerg, Bruzelius, and Rothengatter (2003: 73-151); Flyvbjerg and Cowi (2004); Flyvbjerg (2006, 2008, 2009); and Flyvbjerg, Garbuio, and Lovallo (2009).

[4] Unfortunately there seems to be less and less serious journalism around, which only makes it all the more important to engage with that which is still there.

[5] Others argue that forecasting is done not by planners but mainly by specialized engineers and economists, again letting planners off the hook for bad forecasts. Whereas it is correct that engineers and economists are often involved in forecasting, it is incorrect that planners are not. Planners and their superiors often have the final word in deciding which forecast out of several to use for a specific project, just as planners have been known to tweak forecasts done by engineers and economists until the results suit the needs of project promoters. Our work confirms that of Wachs (1989, 1990), Kain (1990), and Pickrell (1992), who found that planners are actively involved in forecasting (Flyvbjerg 2007a).

[6] Professional organizations are theoretically supposed to use their codes of ethics to penalize and possibly exclude members who do unethical forecasts. Given how widespread unethical forecasts are, it is interesting to note that such penalties seem to be non-existent or very rare, despite codes of conduct that explicitly state that misinforming clients, government, citizens, and others is unacceptable behavior by members of these organizations. This needs to be problematized and debated within the relevant professional organizations. Malpractice in planning should be taken as seriously as malpractice in other professions. To ignore malpractice instead of problematizing and reducing it equals not taking the profession of planning seriously.

[7] Personal communication, author's archives. The American Institute of Certified Planners (AICP) is APA's professional institute, providing certification of professional planners, ethical guidelines, professional development, and more.

[8] A reviewer of the article commented, "for an author so concerned with ethics to point out how he got his way by threatening an editor with unspecified reprisals seems a bit strange." However, there is no reason to be holier than thou and I had no ethical qualms about spelling out the legal implications to the editor, as I saw them. If anyone was acting unethically it was the editor, in trying to drop our agreement. This was a basic negotiation and no code of professional ethics was broken, in my analysis.






---

[9] The one percent does not include the role as media pundit. If this role is included, the time used on media would be substantially higher. However, we typically comment only on things we have actually done research on, declining about 90 percent of invitations to appear in media and thus declining the role of pundit.